\documentclass[aip,graphicx]{revtex4-1}

\usepackage{color}
\usepackage{graphicx}
\usepackage{epstopdf}
\usepackage{amsmath}

\newcommand{\wmk}{~W~m$^{-1}$~K$^{-1}$~}

\begin{document}

\title{Transferability of neural network potentials for varying stoichiometry: phonons and thermal conductivity of Mn$_x$Ge$_y$  compounds}

\author{Claudia Mangold}
\affiliation{Max-Planck-Institut f\"ur Polymerforschung, Ackermannweg 10, 55128 Mainz, Germany}
\author{Shunda Chen}
\affiliation{Department of Chemistry, University of California Davis, One Shields Ave, Davis, 95616, CA, USA}
\author{Giuseppe Barbalinardo}
\affiliation{Department of Chemistry, University of California Davis, One Shields Ave, Davis, 95616, CA, USA}
\author{J\"org Behler}
\affiliation{Universit\"at G\"ottingen, Institut f\"ur Physikalische Chemie, Theoretische Chemie, Tammannstr. 6, 37077 G\"ottingen, Germany}
\author{Pascal Pochet}
\affiliation{Department of Physics, IRIG, Univ. Grenoble Alpes and CEA, F-38000 Grenoble, France}
\author{Konstantinos Termentzidis}
\affiliation{Universit$\acute{e}$ Claude Bernard Lyon 1, CNRS, INSA-Lyon, CETHIL UMR5008, F-69621, Villeurbanne, France}
\author{Yang Han}
\affiliation{Universit\'{e} de Lorraine, CNRS, LEMTA, Nancy ,F-54500, France}
\author{Laurent Chaput}
\affiliation{Universit\'{e} de Lorraine, CNRS, LEMTA, Nancy ,F-54500, France}
\author{David Lacroix}
\affiliation{Universit\'{e} de Lorraine, CNRS, LEMTA, Nancy ,F-54500, France}
\email{david.lacroix@univ-lorraine.fr}
\author{Davide Donadio}
\affiliation{Department of Chemistry, University of California Davis, One Shields Ave, Davis, 95616, CA, USA}
\email{ddonadio@ucdavis.edu}

\date{\today}


 \begin{abstract}
Germanium manganese compounds exhibit a variety of stable and metastable phases with different  stoichiometry. These materials entail interesting electronic, magnetic and thermal properties both in their bulk form and as heterostructures. Here we develop and validate a transferable machine learning potential, based on the high-dimensional neural network formalism, to enable the study of Mn$_x$Ge$_y$ materials over a wide range of compositions. We show that a neural network potential fitted on a minimal training set reproduces successfully the structural and vibrational properties and the thermal conductivity of systems with different local chemical environments, and it can be used to predict phononic effects in nanoscale heterostructures. 
\end{abstract}

\maketitle

\section{Introduction}
\label{intro}

Machine-learning potentials (MLP) provide a versatile tool to study complex materials with diverse local chemical environment with accuracy comparable to that of the electronic structure calculations to which they are trained, which is usually density functional theory (DFT).\cite{Behler:2016bp} 
A few classes of MLPs can actually achieve such level of accuracy and transferability across various states of matter: Successful examples are neural-network  potentials (NNP),\cite{Behler:2007fe,PhysRevLett.120.143001,PhysRevLett.121.265701}  Gaussian approximation potentials (GAP) with smooth overlap of atomic positions (SOAP),\cite{Bartok:2010fj,Bartok:2013cs} moment tensor potentials\cite{Shapeev:2016kn} and spectral neighbor analysis potentials.\cite{Thompson:2015vl}    
The flexible form and extensive number of parameters of these potentials enable accurate simulations of elemental and binary materials across their phase diagram, including high-pressure phases \cite{Behler:2008ft}, liquids \cite{Khaliullin:2010el,Morawietz:2016js} and glasses \cite{Caro:2018bg,Deringer:2018in}, interfaces\cite{Artrith:2012el} and nanostructures.\cite{Gabardi:2017ja,Quaranta:2017kz}
Besides structural stability and total energy, MLPs allow one to model response functions, such as vibrational spectra \cite{Gastegger:2017cb}, and transport coefficients, e.g.  thermal conductivity \cite{Sosso:2012vc,Campi:2015jm,Li:2020ww}. 

Whereas several works proved the efficacy of MLPs in dealing with diverse chemical environments, so far their performance for large variations of stoichiometry in solids has not been systematically tested.
Here we address the transferability of a NN potential of a binary system over a wide range of compositions. For this purpose we consider Mn$_x$Ge$_y$, which is an interesting material with several stable and metastable polymorphs, and potential applications in electronics, spintronics, and thermoelectric energy conversion\cite{Jamet:2006fp, Arras:2010vy, Tardif:2010it, Arras:2011dx, Spiesser:2011kq}. In particular, among the crystalline phases, MnGe is a fascinating topological materia,l for which it was recently measured a large magneto-thermopower,\cite{Fujishiro:2018fp} but its thermal conductivity is unknown. 
Ultimately it would be desirable to attain a reliable description of nanostructured Mn-doped Germanium materials. Experiments suggested that Mn$_5$Ge$_3$, Mn$_{11}$Ge$_8$ and MnGe play an important role in the formation of Mn-Ge phases precipitated in Ge.
Mn$_5$Ge$_3$ and Mn$_{11}$Ge$_8$ are both stable under standard pressure and temperature conditions and they exhibit structural similarities.\cite{Arras:2010vy,Arras:2011dx} 

As we focus on the vibrational properties and heat transport of these systems, our goal is to fit and test a NNP that reproduces correctly the structure, the phonon dispersion relations and the thermal conductivity of the phases of Mn$_x$Ge$_y$ from pure Ge all the way to Mn$_5$Ge$_3$, including the magnetic materials  MnGe and Mn$_{11}$Ge$_8$, so to enable future studies of growth, structural transformations and heat transport in nanostructured Mn-doped Ge films. 
The training set for the NNP is obtained by accurate DFT calculations of total energies and forces. We validate the accuracy and transferability of the NNP by comparing the structural parameters, e.g. equilibrium density and lattice parameters, and the elastic response to hydrostatic compression against those obtained from the calculations of the equations of states of the different Mn$_x$Ge$_y$ systems by DFT. Phonon dispersion relations are validated against those computed by DFT and the lattice thermal conductivity is compared to that obtained by first-principles anharmonic lattice dynamics and the Boltzmann transport equation.\cite{phono3py}  We prove that NNPs enable the calculation of thermal conductivity at room temperature by equilibrium molecular dynamics (EMD) and the Green-Kubo approach.\cite{Kubo:1966dq,ZWANZIG:1965tp}

\section{Systems and Methods}

\subsection{Mn$_x$Ge$_y$ compounds}

\begin{figure}[bt]
  \includegraphics[width=8.5truecm]{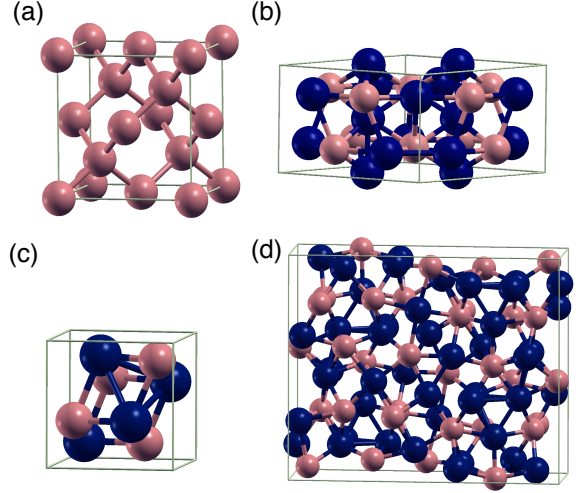}
  \caption{Bulk structures: Ge (a), Mn$_5$Ge$_3$ (b), MnGe (c) and Mn$_{11}$Ge$_8$ (d)}
  \label{pic:bulk_structures}
\end{figure}

The properties of the stable phases of Mn$_x$Ge$_y$ compounds as a function of their composition are thoroughly described by Arras {\sl et al.}~\cite{Arras:2010vy,Arras:2011dx}. The phase diagram is comprised of 16 known phases with composition between pure Ge and pure Mn. Six of them are stable at ambient conditions, while the others are either stabilized by either temperature or pressure and are metastable at ambient temperature and pressure. In the interval of stoichiometric composition of interest to our research, i.e. from Ge to Mn$_5$Ge$_3$, only Mn$_{11}$Ge$_8$ is stable at ambient conditions, while MnGe$_4$, MnGe and Mn$_3$Ge$_5$ are stable at high pressure. In this work we consider the three stable phases, bulk Ge, Mn$_{11}$Ge$_8$ and Mn$_5$Ge$_3 (\eta)$ and metastable MnGe (Figure~\ref{pic:bulk_structures}).
Ge, MnGe and Mn$_5$Ge$_3 (\eta)$ have small unit cells consisting of 2, 8 and 16 atoms, while Mn$_{11}$Ge$_8$ forms a lower-symmetry structure with 76 atoms per cell. We then decided to focus on the first three structure to train and validate the NNP and eventually to test the transferability to Mn$_{11}$Ge$_8$ which was excluded from the training set. Since the NNP used here is constructed as a sum of environment-dependent atomic energies, we expect it to successfully describe also this latter system, because the local chemical bonding environment is quite similar to that of  Mn$_5$Ge$_3 (\eta)$. 

Except for bulk Ge, the Mn$_x$Ge$_y$ phases considered exhibit magnetic ordering at low temperature. Mn$_5$Ge$_3$ is ferromagnetic, as consistently shown both by experiments and DFT calculations.\cite{Arras:2011dx,Arras:2012iva,Forsyth:1999ia} 
The magnetic ordering of MnGe is debated, as early experiments singled it out as antiferromagnetic, while more recent calculations, including the present work, suggest that  ferromagnetic ordering is slightly more stable, by about $0.4$~eV per unit cell.\cite{Arras:2011dx} More recent neutron scattering experiments suggest a chiral magnetic ordering that is beyond the scope of this work to explore further.\cite{Makarova:2012kb}
Mn$_{11}$Ge$_8$ is an anti-ferromagnet at low temperature, which then converts into a ferromagnet above the Neel temperature of 150 K and eventually becomes paramagnetic above the Curie temperature of 274 K.\cite{yamada1990atomic} Our unrestricted spin-polarized calculations confirm  the antiferromagnetic ordering  of Mn$_{11}$Ge$_8$ at zero temperature.

\subsection{First-principles calculations}
\label{sec:abinitio}

We use DFT both to fit and to validate our NN potential. 
DFT calculations, including structural relaxations of unit cells as well as MD runs and single point calculations of structures along MD trajectories,  are performed within the generalized gradient approximation (GGA) by Perdew, Burke and Ernzerhof (PBE) \cite{Perdew:1996iq} as implemented in the plane-wave code Quantum-Espresso \cite{Giannozzi:2009hx}.  
The plane-waves basis set is cut off at 55~Ry and the integration on the first Brillouin zone is carried out on uniform  Monkhorst-Pack meshes of $k$-points\cite{MONKHORST:1976ta} 
 chosen so to ensure that the total energy is converged within $0.005$\,eV/atom for each system. 
The electronic occupation is smeared according to the Marzari-Vanderbilt scheme with a broadening of 0.05~eV, so to achieve more efficient convergence of metallic systems.\cite{Marzari:1999tr}
 The core electrons are described with the projector augmented wave (PAW) method\cite{Blochl:1994uk}. The PAW pseudopotential for Mn was generated using the ATOMPAW code \cite{Holzwarth:2001vk}. We applied an augmentation radius of $1.$~\AA\ to define the region where the ultrasoft pseudopotential is effective. Mn pseudopotentials are set to treat semi-core electrons explicitly, which is essential to provide the correct level degeneracy for the Mn atom in vacuum.\cite{Arras:2011dx}
Some Mn$_x$Ge$_y$ compounds exhibit magnetic behavior, as they are either ferromagnetic, anti-ferromagnetic or non-collinear (chiral), thus,
to account for magnetic ordering, we perform unrestricted spin-polarized calculations.

{\sl Ab initio} phonon dispersion relations are calculated  using either density functional perturbation theory (DFPT) \cite{Baroni:2001tn} or the frozen phonon approach, in which the force constants are computed by finite differences over atoms displacements in a sufficiently large supercell. 
Given the short-range nature of the forces in the systems considered, both methods provide results with comparable accuracy, as we verified for bulk Ge.  
While DFPT is more general, it becomes more computationally expensive for crystals with large number of atoms in the unit cell. Hence for Mn$_5$Ge$_3$ and Mn$_{11}$Ge$_8$ we employ the frozen phonons approach with displacements of the atoms of 0.01 \AA. 

Thermal conductivity is calculated using \emph{phono3py}\cite{phono3py}, which implements the solution to the linearized BTE in the relaxation time approximation. Second and third order interatomic force constants (IFC) are computed by DFT using the PAW method as implemented in the VASP code\cite{VASP1,VASP2,VASP3,VASP4,VASP5,VASP6}. To obtain forces, total energies are converged with an accuracy better than $10^{-8}$eV in supercells whose size is given in Tab. \ref{table:database}. 

The BTE is then solved in the relaxation time approximation, which allows one to express the lattice thermal conductivity as
\begin{equation}
\kappa=\frac{1}{VN_q}\sum_{i,q} c_i(q)\textrm{v}_i(q)\otimes\textrm{v}_i(q)\tau_i(q)
\label{eq:kappa}
\end{equation}
 $\textrm{v}_i(q)$ are the phonon group velocities obtained as $\frac{\partial \omega_i(q)}{\partial q}$ and $c_i(q)$ is the phonon heat capacity, i.e. $c_i(q)=\hbar \omega_i(q) \frac{\partial n(\omega,T)}{\partial T}$ with $n$ being the Bose-Einstein distribution function. 
 $\omega_i(q)$ and $\textrm{v}_i(q)$ can be obtained from the harmonic force constants alone, but to compute $\tau_i(q)$, the third order anharmonic force constants are required.\cite{phono3py}

\subsection{Neural Network Potential: Details and Training}

We employ the Behler-Parrinello NNP scheme\cite{Behler:2007fe,Behler:2014jc,doi:10.1002/anie.201703114} to generate a transferable linear-scaling MLP for germanium and Ge-Mn systems. This general neural network scheme consists of a set of symmetry functions\cite{Behler:2011it} that feed the atomic coordinates to a number of hidden layers, which provide an analytical expression of the energy. Forces are obtained as the analytical negative gradient of the energy function. For our systems we chose a relatively simple architecture using the code RuNNer.\cite{Behler:2015gv,doi:10.1002/anie.201703114} It consists of two hidden layers, each containing 20 nodes. For each node a hyperbolic tangent is used as non-linear ``activation function", while the identity $f(x)=x$ is used as activation function for the output layer. 
48 atom-centered symmetry functions represent chemical environment of each atom up to a cutoff $r_c=$6.35 \AA. A cutoff function is defined so that the potential goes to zero with continuous first and second derivatives at $r_c$.\cite{Behler:2011it} 
The NNP constructed in this way is short-range, hence linear scaling. As it neglects long-range it may not be suitable to treat strongly ionic systems, but this is not the case for Mn-Ge. 
A similar approach was employed to fit a NNP for GeTe,\cite{Sosso:2012hi} showing excellent performance in describing the structure of its liquid and amorphous phases and the crystallization mechanism,\cite{Sosso:2013ik,Gabardi:2017ja}, as well as the structure and vibrational properties of nanowires.\cite{Bosoni:2019cq}
\textcolor{black}{Whereas in principle it would be possible to use larger NN, with more layers or more nodes per layer, that would mean adding even more parameters to fit, thus making it very difficult to achieve accuracy with a limited training set.}

\begin{figure}[bt]
  \includegraphics[width=8.5truecm]{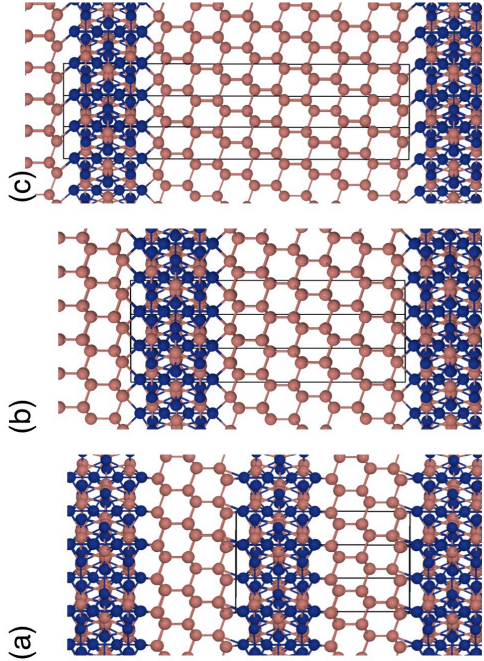}
  \caption{Ge[111]/Mn$_5$Ge$_3$[001] heterostructures with varying thickness of the Ge layer: namely, $7$\,\AA \,(a), $18$\,\AA \,(b), $28$\,\AA \,(c) (black boxes correspond to the respective unit cells).}
   \label{pic:different_interfaces}
\end{figure}

Fitting a reliable NNP requires a comprehensive training database.  As we are addressing a range of stoichiometry from pure Ge to Mn$_5$Ge$_3$, besides these two compositions we include in the database also crystalline MnGe as a system with intermediate stoichiometry. Furthermore, to extend the applicability to nanostructured materials, e.g. superlattices Ge/Mn$_x$Ge$_y$ and Mn$_5$Ge$_3$ nanoinclusions in bulk Ge, we add to the training set a superlattice that features Ge$[ 111 ]$/Mn$_5$Ge$_3 [001]$  interfaces. 
This superlattice consists of a thin Ge [111] slab ($\sim 7$~\AA ) and an even thinner Mn$_5$Ge$_3$ slab oriented in the [001] direction. These two surfaces form a clean interface with good lattice matching ($\sim 6\%$), which was observed in epitaxial growth experiments.\cite{Zeng:2004im,OliveMendez:2008kd,Xie:2018hq}. The resulting interfacial structures contains 44 atoms in its unit cell (structure (a) in Fig. \ref{pic:different_interfaces}, while the larger models represented in panels (b) and (c) are used for calculations only). These structures feature low-energy and low-strain interfaces between Ge and Mn$_5$Ge$_3$: the features of these interfaces favor phase separation of Mn-Ge solid solutions leading to the formation of nanostructured films.\cite{Arras:2012iva}

To generate a sufficient number of training structures we performed Born-Oppenheimer Molecular Dynamics (BOMD) simulations of supercells containing about 100 atoms obtained replicating the crystalline structures.  BOMD simulations in the constant volume canonical (NVT) ensemble are performed for systems at various densities and temperatures between 300 and 700 K. We stress that it is important to use systems at different densities to train the NNP on larger variations of bond lengths and angles. 
In particular we perform MD simulations for systems with lattice parameters up to  $\pm 4$~\% of the equilibrium lattice constants.
In these simulations the electronic structure is computed at the $\Gamma$-point only,  and the Newton's equations of motion are integrated with a timestep of 4 fs. 
This setup is sufficient to sample the configurational space of the system. 
From the BOMD trajectories we extract statistically uncorrelated frames {-- approximately one every 100 fs of MD trajectory, with a randomized time lag --}, for which we perform well-converged electronic structure calculations, as described in the previous section.
These calculations provide accurate total energies and forces that are both used to train the NNP. 
{Since we consider very different models, with different numbers of atoms of the two types, total energies would not be comparable across the training set. For this reason, we fit the NNP on atomization energies, which have a well-defined physical meaning and are consistent for different systems.} 
Further training configurations are produced while testing the NNP in MD runs, when configurations occur that are out of the interpolation range of the NNP, so to add cyclical self-consistence to the training procedure. 
\begin{table}
  \caption{Data base of structures used to train and test the neural network potential. While we generated an extensive database of structures, we tried to find a minimal training set, so to avoid over-fitting problems and reduce the number of correlated structures.}
  \label{table:database}
    \begin{tabular}{lcccc}
      \hline\noalign{\smallskip}
       Species                                      &  cell       & Supercell (\#\,atoms)             &  \#\,structures & training set \\
      \noalign{\smallskip}\hline\noalign{\smallskip}
        Ge                                             & fcc           &  $2\times2\times2$ ($64$)    & 2522  &  $1340$\\
        Mn$_5$Ge$_3$                        & hex.    &  $1\times1\times2$ ($32$)   &  2396  &  $731$  \\
        MnGe                                        & sc            &   $2\times2\times2$ ($64$)   & 1973  &  $640$  \\
        Ge/Mn$_5$Ge$_3$  & hex.  &   $1\times1\times1$ ($44$)                     &  4794 &  $650$  \\
        \noalign{\smallskip}\hline\noalign{\smallskip}
                                                          &                      &                                   Total:  &  11685 & $3361$ \\
    \noalign{\smallskip}\hline
   \end{tabular}
\end{table}
\begin{figure}[bt]
  \includegraphics[width=8.5truecm]{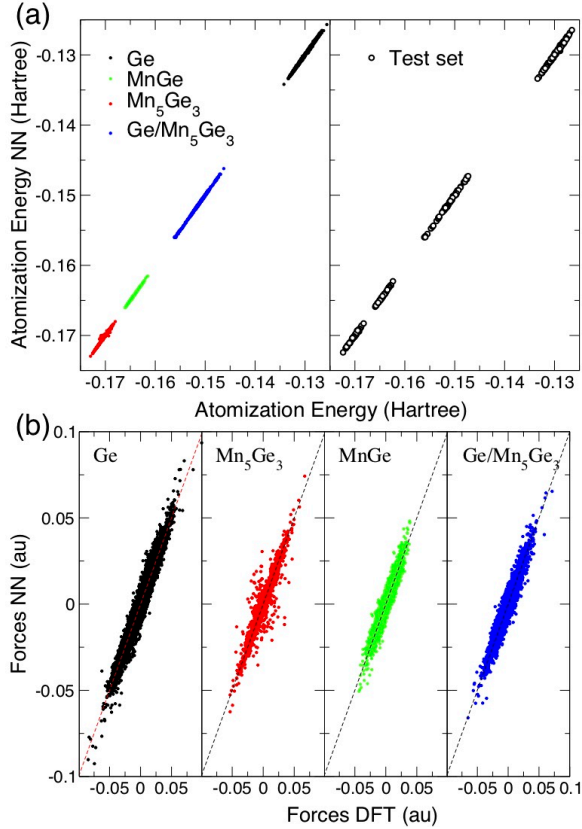}
  \caption{Energies (a) and forces (b) computed by the neural network potential against the DFT reference. In panel (a) energies are shown for the training (left) and test sets (right). In panel (b) forces are shown for the test set only and are given for each separate system. Both energies and forces are in atomic units, Hartree and Hartree/Bohr, respectively.}
   \label{pic:NNforces}
\end{figure}

Since for each frame the number of force components, corresponding to the number of degrees of freedom of the system, overwhelms the single total energy entry, only a randomly chosen fraction of the force components is used to fit the NNP. The best fit is obtained a fraction of the forces corresponding to $\sim 3$ times the number of total energies. 
{The parameter optimization is carried out using an adaptive Kalman filter.\cite{Witkoskie:2004be} With this optimization algorithm the NN parameters are updated upon the presentation of each individual energy or force component, so that a global loss function, which would combine errors from the energies and the forces, is not required. To obtain a balanced statistical weight of the input data, we define an error threshold on energies and forces, and we update the NN parameters only when the error exceeds the threshold. In these way the optimization process selects and gives more weight to the configurations that are less accurately represented. The error threshold itself is not fixed but specified with respect to the current RMSE such that the threshold decreases along with the improvement of the fit.}
\textcolor{black}{To avoid overfitting, we implemented an iterative search of structures that are not well represented, that is those for which different NNs trained to the same data set predict very different energies and forces. This search is carried out iteratively while improving the potential until convergence is reached. In addition we continuously apply the early stopping method, i.e. not all available training points are used for optimizing the NN parameters, but a part of the data set is kept as an independent test or validation set to assess the quality of predictions for new structures.\cite{Behler:2015gv}}

Details of the training set used to generate the NNP are given in table~\ref{table:database}.
This setup leads to a root mean square error (RMSE)  of $2.2$~meV/atom on the energies of the training set and $2.5$~meV/atom for the test set, and it gives a RMSE of $0.085$~eV/\AA\ and $0.089$~eV/\AA\  for the forces on the training and test sets, respectively. {Energies per atom obtained with the NNP against the DFT reference are shown in Figure~\ref{pic:NNforces}a for both the training and the test set.
Figure~\ref{pic:NNforces}b shows that the error on the forces is nearly equivalent for all systems: although slightly larger deviations from the DFT reference occur for Mn$_5$Ge$_3$, the RMSE on the forces is actually similar for the four structures considered.} The parameters of the NNP are available as data files in the supporting information (SI). \textcolor{black}{These are plain ASCII files, in a format readable to the code RuNNer, and consist of: a commented input file for RuNNer, the weights of the nodes of the NN layers and the scaling factors for normalizing the range of the symmetry functions.}

Former works adopted very large databases of structures for the NNP training\cite{Artrith:2011em,Artrith:2012fw,Sosso:2012hi,Morawietz:2013cc}: for example the NNP for GeTe \cite{Sosso:2012hi}, which has the same symmetry functions and similar network structure as ours, was fitted for more than $30\,000$ structures.
Here, however, we try to find a minimal database with about one tenth of the structures. Whereas on one side we have a range of compositions, on the other side we can focus for the moment on crystalline structures and superlattices, thus limiting the need for transferability to an extremely broad range of chemical environments. 
Therefore we started the refinement process with MD simulations with a preliminary NN potential based on only several hundreds of structures per Mn-Ge phase. Problematic structures from these MD runs were picked and added to the training set in order to systematically improve the NNP. 

\subsection{Thermal Conductivity from Molecular Dynamics Simulations}

The fitted NNP is exploited to perform MD simulations and to compute the lattice thermal conductivity from the fluctuation of the heat current at equilibrium using the Green-Kubo expression for transport coefficients:
\begin{equation}\label{eq.kubo}
\kappa_{\alpha\beta}=\dfrac{V}{k_BT^{2}}\int\limits_{0}^{\infty}\langle J_\alpha(0) J_\beta(t) \rangle dt
\end{equation}
where $\kappa_{\alpha\beta}$ ($\alpha\beta$=\textit{x},\textit{y},\textit{z}) is the ${\alpha\beta}^{th}$ component of the thermal conductivity tensor,
$V$ is the volume of the simulation cell, $k_B$ is the Boltzmann constant, $T$ is temperature, and $J_\alpha$ is the $\alpha^{th}$ component of the heat current vector.

The heat current consists of the sum of a kinetic and a potential energy term: 
\begin{equation}
 \mathbf{J}=\mathbf{J}_{kin}+\mathbf{J}_{pot} = \sum_i E_i \mathbf{v}_i + \sum_i \mathbf{r}_i \frac{dE_i}{dt}
\end{equation}
The NNP is a short-range analytical function of the coordinates of the system, and the use of atom centered symmetry functions are chosen so that the total energy of the system is expressed as the sum of atomic contributions: $E_{tot}=\sum_i E_i$. This observation is sufficient for us to deal with the kinetic term, which usually provides a negligible contribution to the total thermal conductivity of solids. 
Furthermore, the symmetry functions amount to pair ($f(r_{ij})$) and three-body functions ($f(r_{ij},r_{ik},r_{jk})$), where $r_{ij}$ indicates the vector connecting two atoms within the chosen cutoff $r_c$.\cite{Behler:2016bp} 
These features of the NNP allow us to define a pairwise force $\mathbf{F}_{ij}$ between two atoms $i$ and $j$, which includes all the three-body contributions. This pairwise force is defined so to satisfy two conditions: The total force on atom $i$ is given by $\mathbf{F}_i=\sum_{j\ne i}\mathbf{F}_{ij}$ and Newton's third law holds, i.e. $\mathbf{F}_{ij} = - \mathbf{F}_{ji}$. These conditions allow us to define a pairwise atomic stress tensor $\sigma_{i}= -\frac{1}{2}\sum_{j\ne i}\mathbf{r}_{ij}\otimes \mathbf{F}_{ij}$, which, in turn, can be used to calculate the heat current as:
\begin{equation}
 \mathbf{J}_{pot} = \sum_i \sigma_i \cdot \mathbf{v}_i
\end{equation}
The details on how to consistently derive the two-body force and the heat current expression for multi-body potentials are provided in Ref.~\cite{Fan:2015ba}

The MD simulations were conducted with a modified version of DLPOLY v2.19\cite{todorov2006dl_poly_3} interfaced with RuNNer, which comprises the calculation of the heat current from the decomposition of the many-body NNP in local energy density.\cite{Sosso:2012vc,Fan:2015ba} The same approach and software was employed to characterize thermal transport in phase change material GeTe\cite{Sosso:2012vc,Campi:2015jm}. We considered domain sizes up to 6$\times$6$\times$6 cubic conventional cells for Ge and MnGe, and 5$\times$5$\times$6 unit cells for solid Mn$_5$Ge$_3$. 
Periodic boundary conditions were applied in $x$, $y$, and $z$ directions. 
The Verlet algorithm is used to integrate the equations of motion with 1~fs time step.  In all simulations, the atomic systems were first equilibrated in a $NPT$ (constant: number of atoms, pressure, and temperature) ensemble for 1~ns before being switched to a $NVE$ (constant: number of atoms, volume, and energy) ensemble for another 1~ns. Berendsen barostat and thermostat\cite{berendsen1984molecular,morishita2000fluctuation} were used to control the pressure and the temperature of the systems during equilibration runs.
 Then, the following 10~ns simulation in $NVE$ ensemble was taken for data production. The flux fluctuations are computed each 1~fs and the integral is sampled over 1000 values. The correlation time upper limit of our calculations was chosen to be 50~ps. Each simulation was run for 20~times with independent initial velocity distributions. It is an inherent assumption in this study that 20 independent simulations provide a representative sample for the relevant statistical analysis. Finally, we reported the average of the 20 independent MD runs as the predicted thermal conductivity and  the standard error as its uncertainty.

\section{Results and Discussion}
\subsection{Structural Parameters and Equation of State} 

\begin{table}
    \caption{Lattice parameters and bulk moduli of Mn-Ge bulk materials evaluated with the NN potential and with \textit{ab initio} DFT calculations.}
    \label{table:bulk}
    \begin{tabular}{ l l  l l | c c c c }
      \hline 
       Species         & \multicolumn{3}{c}{ a [\AA] and c/a}\vline  & \multicolumn{3}{c}{$B_0$\,[GPa]} \\
        \hline\hline
                       & NN     &   DFT  &   Exp.  & NN & DFT & Exp. \\
                       \hline
       Ge              & $5.75$ &  $5.76$ & $5.66$  & $62.3$ &  $60.1$ & $76.8$\\
         \hline 
       MnGe            & $4.75$ &  $4.74$ & $4.80$ & $113.2$ &  $114.2$ &    -- \\
      \hline 
       Mn$_5$Ge$_3$   & 7.15      & 7.14     & 7.18     & $104.2$\  &  $111.0$\  & $110$\  \\
                                    &  0.697   &  0.697  &  0.703  &               &                &  \\
    \hline\hline
   \end{tabular}
\end{table}
\begin{figure}
  \includegraphics[width=8.5truecm]{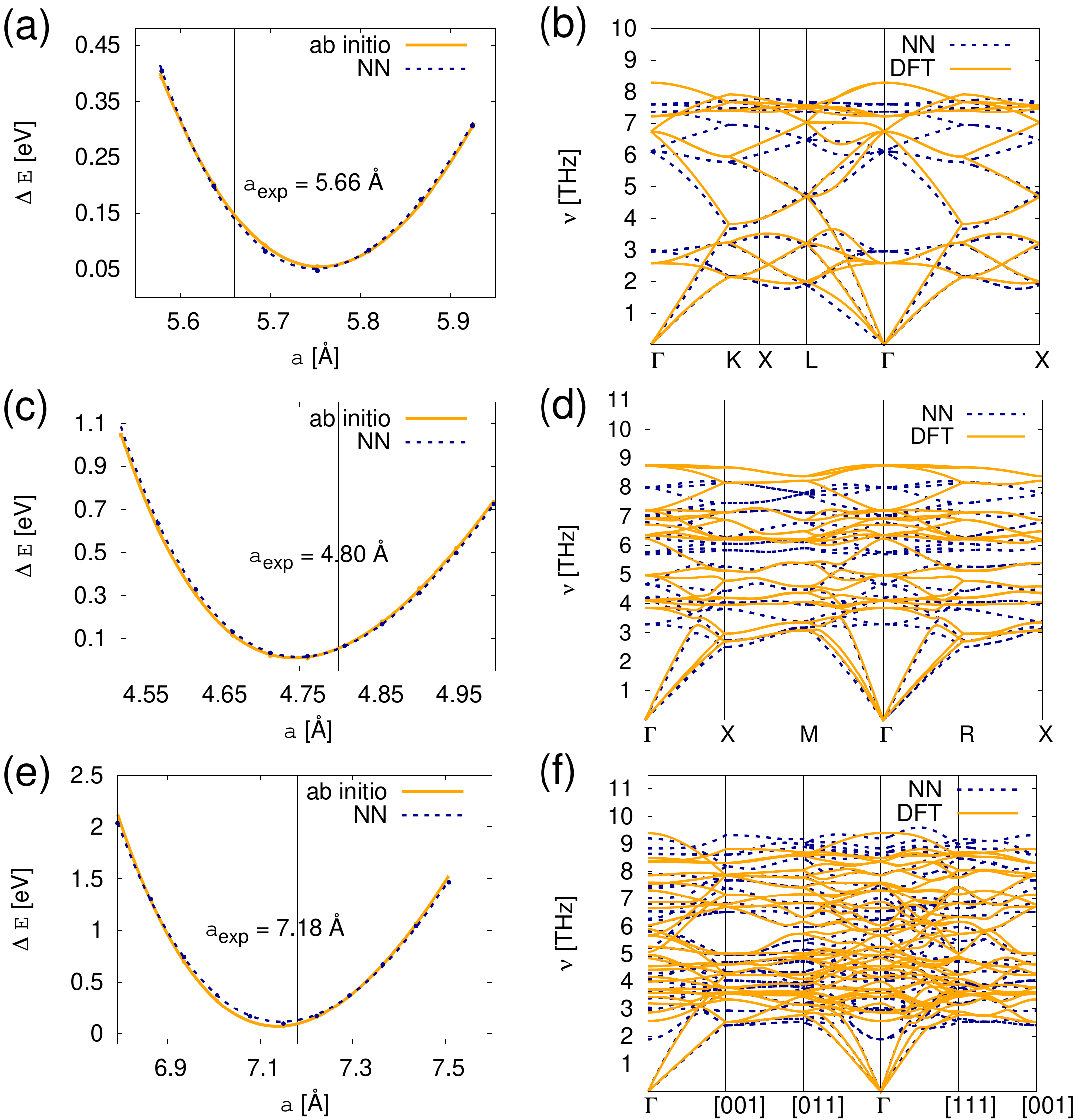}
\caption{Murnaghan fit of the equation of state and phonon dispersion relations of Ge (a,b), Mn$_{5}$Ge$_{3}$ (c,d) and MnGe (e,f) : Comparison of results from \textit{ab initio} (orange solid line) and the neural network potential (black dashed line).}
   \label{pic:eos}
\end{figure}

We first verified that the fitted NNP reproduces the structural and vibrational properties computed by DFT of the systems used to build the training set.
In Table~\ref{table:bulk} we compare the lattice parameters and the bulk moduli ($B_0$ of Ge, MnGe and Mn$_5$Ge$_3$ obtained by computing the equations of state by NNP, DFT and experiments. Equilibrium lattice parameters and $B_0$ are obtained by fitting the equation of state (EOS) to a Murnaghan function~\cite{Murnaghan244} (see Figure~\ref{pic:eos}).
DFT results are in good agreement with experiments confirming that the adopted computational framework is reasonable to model these complex materials.\cite{Arras:2010vy}  
The agreement between DFT and NNP lattice parameters is excellent to 0.01 \AA\ for the lattice parameters. In addition NNP also provides bulk moduli in good agreement with DFT, differing at most by 6$\%$ for Mn$_5$Ge$_3$, thus suggesting that the NNP reproduces well elastic deformations, which are intrinsically connected to acoustic phonons.

\subsection{Phonons} 

Producing reliable phonon dispersion relations is the first essential step for an empirical potential to predict the lattice thermal conductivity of a material. 
We computed the phonon dispersion relations of bulk Ge, Mn$_5$Ge$_3$ and MnGe along a high symmetry path in the first Brillouin zone. The interatomic force constants to construct the dynamical matrix were calculated using the finite-differences supercell approach, as discussed in section~\ref{sec:abinitio}. This approach is justified by the short-range nature of the forces in the systems considered, however we were careful to verify the convergence of the DFT dispersion relations as a function of the size of the supercell used. 
As the NNP is short-range by construction, the calculation of the phonon dispersion relations does not suffer from size convergence issues, as long as the supercell is twice as large as the interaction cutoff radius of the NNP. 
NNP and DFT phonon dispersion relations,  shown in Figure~\ref{pic:eos}(b,d,f), are in very good agreement, especially for what concerns the acoustic branches, which provide the main contribution to heat transport. Minor discrepancies occur at higher frequency for the optical branches. 

\begin{figure}
 \includegraphics[width=7.truecm]{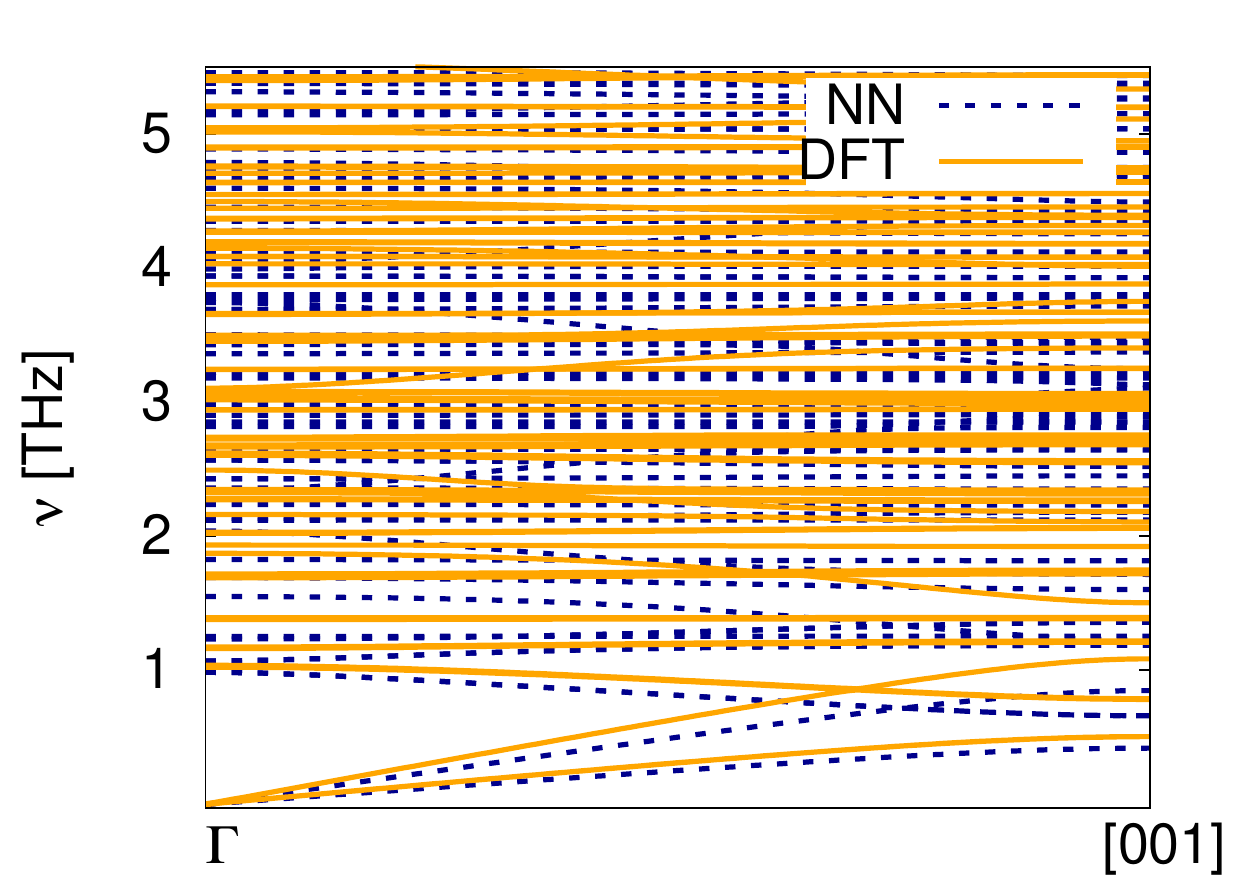} 
   \caption{Phonon dispersion relation of the interface Ge[111]/Mn$_{5}$Ge$_{3}$[001]: Comparison of results achieved with \textit{ab initio} DFT and the NN potential calculations.}
   \label{pic:interface_phonons}
\end{figure}

Finally, we verified that the phonons computed by NNP and DFT agree also for the Ge$[ 111 ]$/Mn$_5$Ge$_3 [001]$ superlattice used in the fitting procedure. The dispersion relations for frequencies below 5.5 THz across the superlattice planes are shown in figure~\ref{pic:interface_phonons}. Agreement between DFT and NN calculation remains satisfactory, although the longitudinal acoustic (LA) mode is slightly softened with the NNP with respect to DFT.

\subsection{Transferability of the NN potential}

Neural networks are in general a powerful approach to interpolate complex data sets but they are not reliable when it comes to extrapolation. However, NNPs may turn out transferable to phases that were not included in the training set, provided that such phases share a similar local chemical environment as the ones used for training.\cite{Behler:2008ft} In this Section we test the transferability of our Mn$_x$Ge$_y$ NNP to the Mn$_{11}$Ge$_8$ phase and to Ge/Mn$_5$Ge$_3$ heterostructures with different superlattice spacing (Fig.~\ref{pic:different_interfaces} (b) (c)). 

\begin{figure}
   \includegraphics[width=7.5cm]{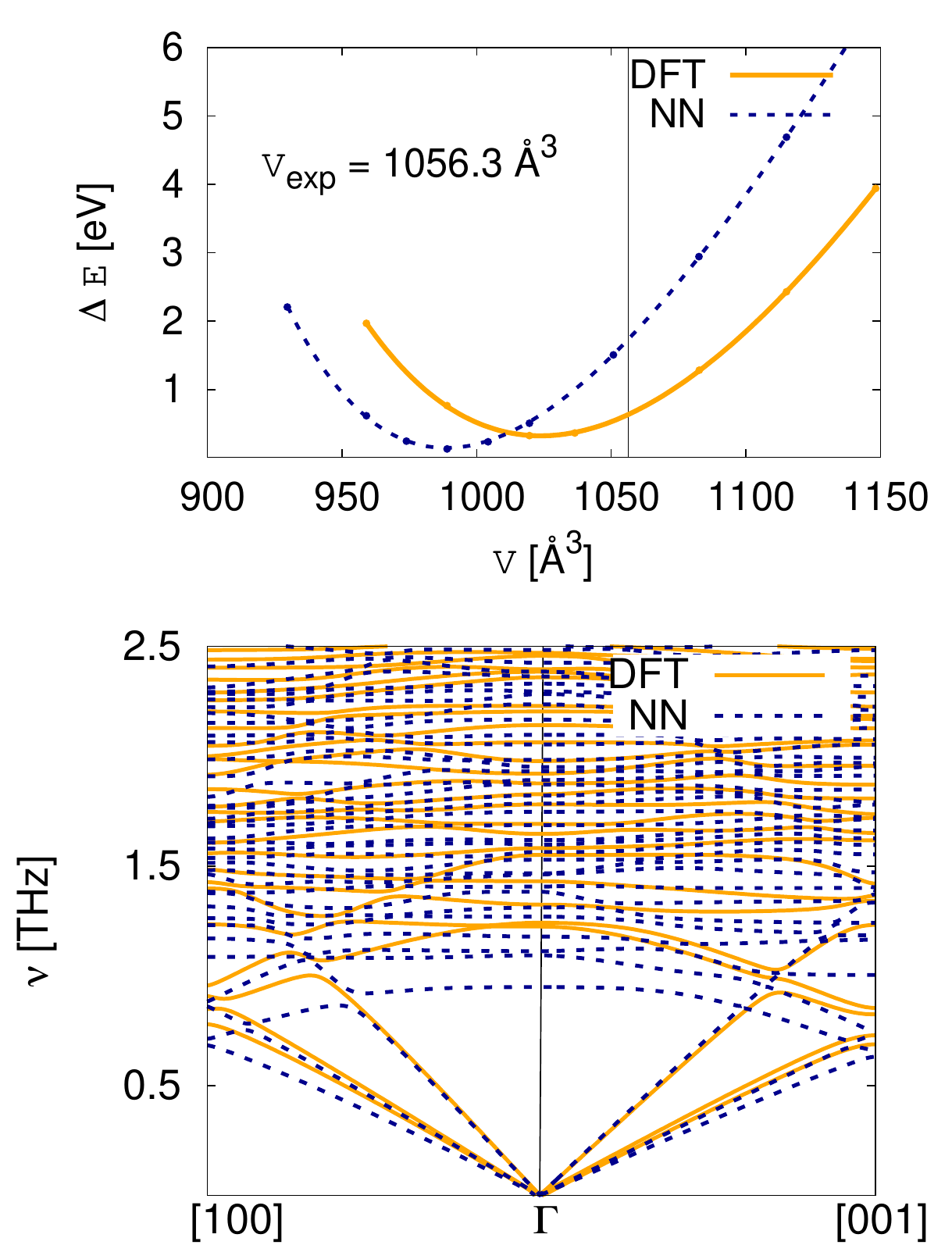}
   \caption{Equation of state (total energy of the unit cell vs. cell volume) (top panel) and phonon dispersion relation (bottom panel) of Mn$_{11}$Ge$_8$: Comparison of results from \textit{ab initio} and the NN potential.}
   \label{pic:mn11ge8_phonons}
\end{figure}

\paragraph{Mn$_{11}$Ge$_8$} is not part of the training set of the generated NNP, but, since it entails structural similarities to Mn$_5$Ge$_3$, it is reasonable to expect that the NNP would perform well in reproducing its structural and vibrational properties. 
 We evaluate the Murnaghan EOS for Mn$_{11}$Ge$_8$ (see Fig. \ref{pic:mn11ge8_phonons} (a)) and find it in good agreement with ab initio and experimental results: The deviation in volume is only $\sim 2\%$ compared to experiment \cite{Ohba:a23234} and less than $1.5\%$ compared to former \textit{ab initio} calculations.\cite{Arras:2011dx}
The error in the bulk modulus is much larger, however acceptable. The bulk modulus obtained with NNP is 142 GPa to be compared with 105 GPa computed by DFT.
Although the NNP reproduces the structure of Mn$_{11}$Ge$_8$ with significantly less accuracy than for the phases included in the training set, the overall agreement with DFT is fairly good.  

We further use the NNP to compute the phonon dispersion relation using a $1\times2\times1$ supercell, consisting of 152 atoms. 
The NNP dynamical matrix is computed by finite differences: Although the supercell is fairly large, the same approach can be used to compute the phonons at the DFT level. 
The comparison between NNP and DFT phonon dispersion relations in the $\Gamma-X$ and $\Gamma-Z$ directions is shown in Fig. \ref{pic:mn11ge8_phonons}(b), zooming into the frequencies below 2.5 THz for the sake of clarity. 
The transverse acoustic (TA) phonon modes evaluated with the NN potential have slightly lower frequencies and lower group velocity (i.e. the slope of the dispersion curve) than the corresponding modes from DFT, whereas the LA mode is reproduced very well by the NNP, especially at center zone. 
In turn, the low-frequency optical modes exhibit a significant shift to lower frequencies, which affects the dispersion of LA toward the edge of the Brillouin Zone. 

\subsubsection{Ge/Mn$_5$Ge$_3$ Superlattices}
NNP transferability has also been tested for the Ge/Mn$_5$Ge$_3$ superlattices by varying the spacing between Mn$_5$Ge$_3$ layers. The interface modeled from the Ge[111] and the Mn$_5$Ge$_3$[001] surface exhibits a thin Ge layer (only $\sim 7\,$\AA\,) and its hexagonal unit cell contains as few as 44 atoms (with only five layers of Ge atoms in between two Mn-Ge phases).  However, in the present research project we aim at a description of Mn-Ge structural features like nano-columns or -clusters in a surrounding Ge matrix, i.e. much larger amount of Ge compared to Mn-Ge phase. Therefore we stepwise increased the Ge layer in the interfacial structure of Ge[111]/Mn$_5$Ge$_3$[001] reaching Ge layers of $\sim 18\,$\AA\, and $\sim 28\,$\AA\, thickness respectively (see Fig. \ref{pic:different_interfaces}). These heterostructures (comprising $62$ and $80$ atoms in the respective unit cell) were not part of the training set of the NNP. Yet we expect NNP to provide a reliable representation of these systems, since the structure of the interfaces  does not change with an increase of the Ge layer thickness.

\begin{figure}
   \includegraphics[width=8.5cm]{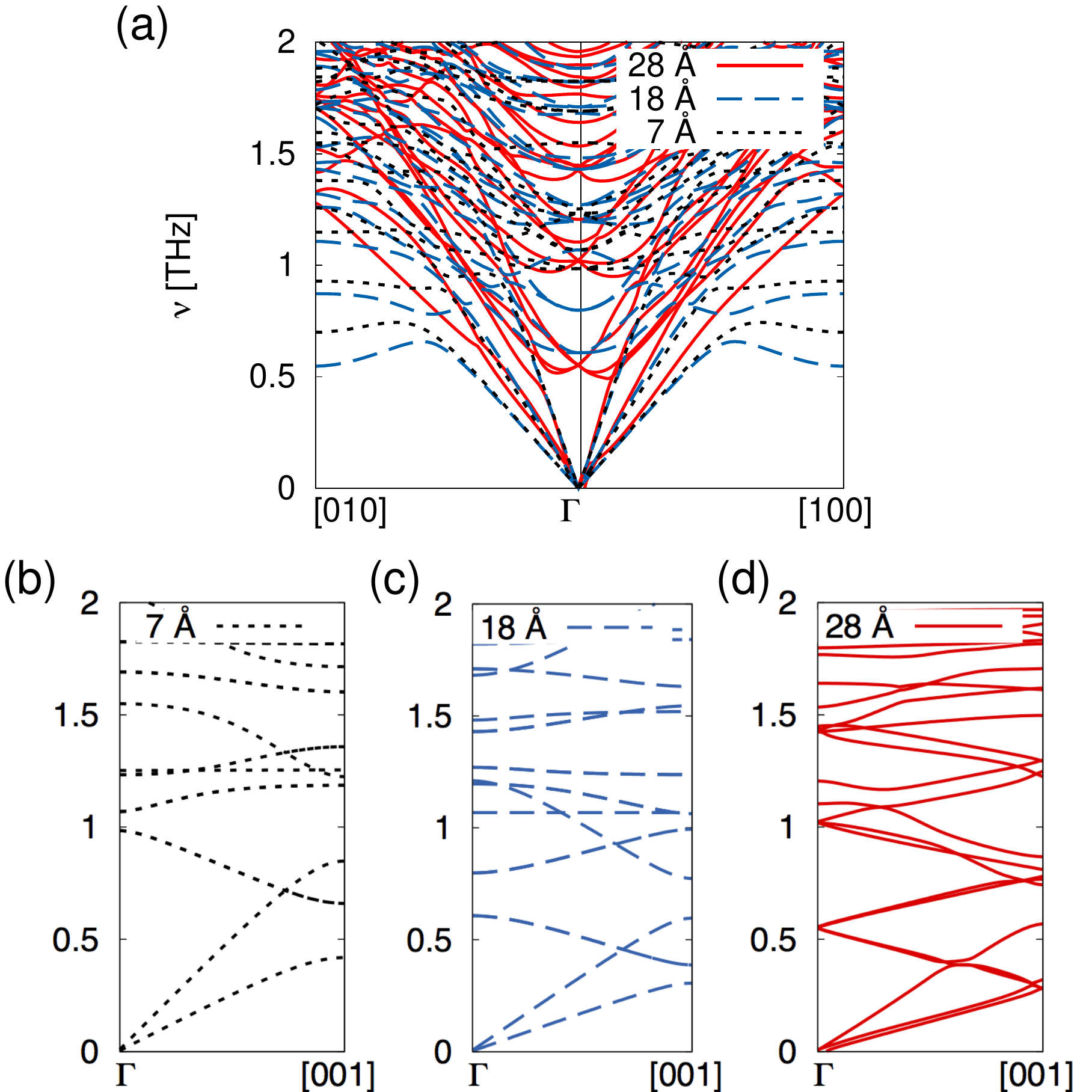}
     \caption{In-plane (a) and cross-plane (b,c,d) phonon dispersion relation of three Ge[111]/Mn$_{5}$Ge$_{3}$[001] interfaces with  of the Ge-layer thickness of 7, 18 and 28 \AA, computed using the neural network potential.}
   \label{pic:phonSL}
\end{figure}

In Figure~\ref{pic:phonSL} we show the phonon dispersion relations of all three interfacial structures, \textcolor{black}{computed using NNP}: The {top} panel provides a close up look into the low-frequency in-plane modes for the three structures, which have the same in-plane cell parameters. The slope of the TA modes depends on the thickness of the Ge layer in a non-monotonic way, with the system with the thickest Ge layer entailing the steepest TA slopes and the highest frequencies at zone boundary. The LA mode near the $Gamma$ point is much less affected by the Ge layer, indicating that the systems have the same longitudinal speed of sound. 
The cross-plane cells have different lattice parameters, thus it makes no sense to compare the cross-plane [001] dispersion relations on the same graph: we show the low-frequency  phonon dispersion relations in the  three  separate bottom panels in Figure~\ref{pic:phonSL}.  
For the system with the thinnest Ge layer (7~\AA ) the dispersion curves in the [001] direction exhibit a gap at 1 THz, which shifts at lower frequency and widens in the system with 18~\AA\ Ge layer. The gap vanishes when the Ge layer is 28~\AA. 
\textcolor{black}{Calculations of heterostructures with even larger lattice parameter are made accessible by the use of NNPs at a frugal computational cost.}
These findings indicate that NNPs can also be used to efficiently design phononic structures, such as superlattices, including the non-trivial chemical features of the interfaces within these heterostructures.

\subsection{Molecular Dynamics and Thermal conductivity}

In this section we present the results of the thermal conductivity for the three Mn$_x$Ge$_y$ crystals used to fit the NNP, as obtained by equilibrium MD and the Green-Kubo approach.
MD runs on supercells of Ge, Mn$_{5}$Ge$_{3}$ and MnGe with several hundreds of atoms showed a stable behavior at 300\,K. Ge as well as MnGe supercell also showed stable MD runs over several ps also at elevated temperatures (500/700\,K). We only observed rare instabilities for Mn$_{5}$Ge$_{3}$ at 700 K, indicating that further high-temperature structures should be added to the NN training set, if simulations under these conditions become necessary. These simulations scale linearly with the number of atoms ($N$) in the simulation cell, as opposed to DFT that scales like $N^2log(N)$, thus NNP can be used to compute the thermal conductivity by MD, testing size convergence and performing a sufficiently large number of runs to achieve a good statistical accuracy.

\begin{center}
\begin{figure*}[h]
\includegraphics[width=\columnwidth]{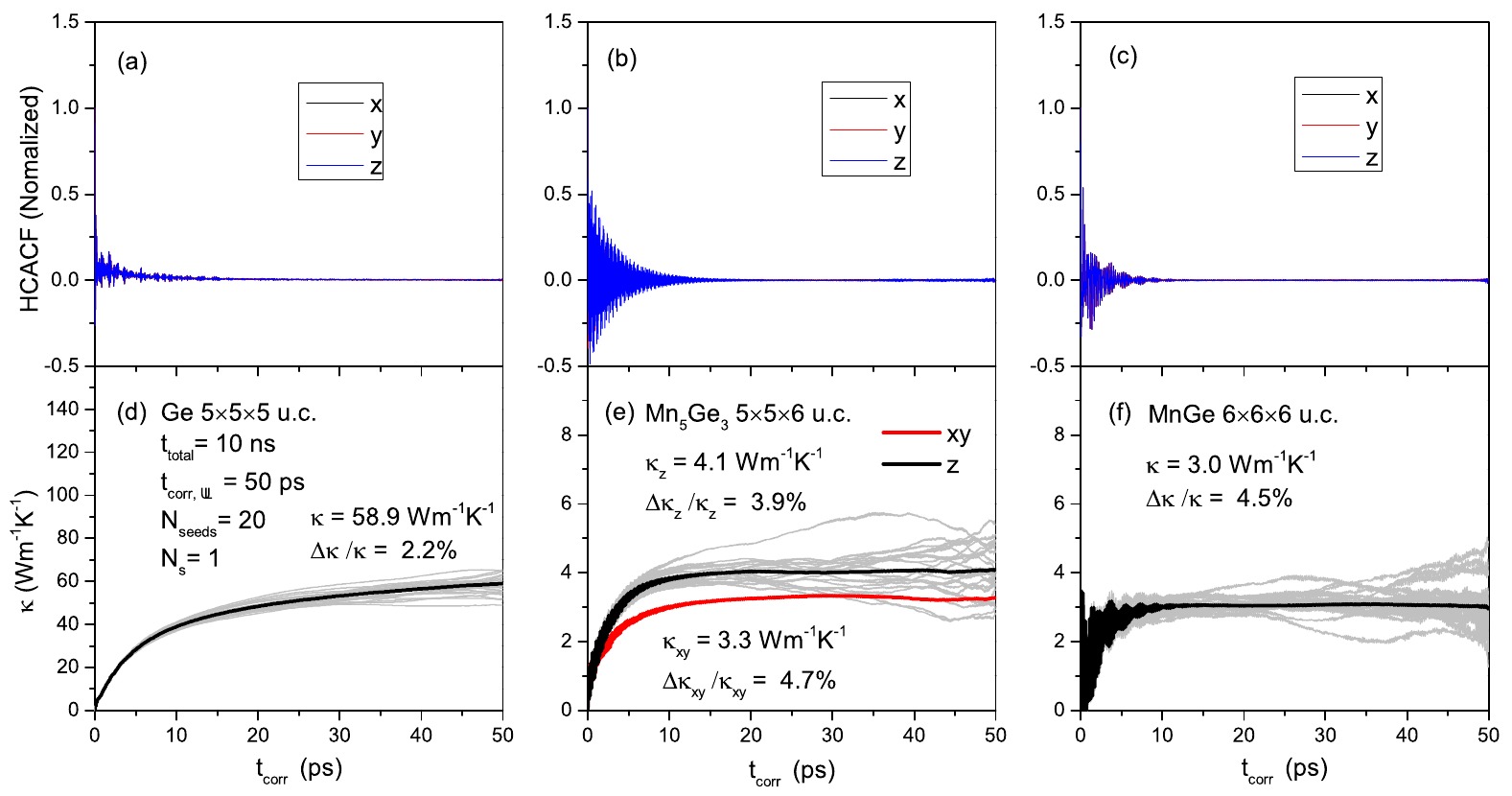}
\caption{\label{fig.emd} Normalized heat current auto-correlation function (HCACF) of single MD runs for Ge (panel a), Mn$_5$Ge$_3$ (b) and MnGe (c). (d)-(f) Running integrals used to estimate the thermal conductivity obtained by averaging over 20 independent simulations for the same three compounds. The light-grey curves represent the thermal conductivities in the \textit{z} direction for each of the 20 independent simulations and the superimposed thick curves represent the corresponding average values. The obtained average thermal conductivities $\kappa$ in each direction and the corresponding standard errors $\Delta\kappa$ are also indicated in each panel along with the key details of the simulations. Since Mn$_5$Ge$_3$ is anisotropic in panel (e) the \textit{x/y} value of the thermal conductivity is represented in red.
}
\end{figure*}
\end{center}

Figure~\ref{fig.emd} displays the normalized heat current autocorrelation function (HCACF) for one example run, the running integrals and their average over 20 statistically independent runs for Ge (supercell 5$\times$5$\times$5, 1000 atoms), Mn$_5$Ge$_3$ (5$\times$5$\times$6 supercell, 2400 atoms) and MnGe (6$\times$6$\times$6 supercell, 1728 atoms). 
All the HCACF (Fig.~\ref{fig.emd} (a-c)) decay rapidly to zero, thus making the evaluation of the integral in Equation~\ref{eq.kubo} relatively straightforward, with a small uncertainty on the final estimate of the thermal conductivity upon averaging. We note that the HCACFs of Mn$_5$Ge$_3$ and MnGe exhibit large fluctuations in the short time scale. These fluctuations, which are absent for pure Ge, are a signature of the mass difference between the elements in binary compounds, as it was formerly seen for the doped clathrate Sr$_{6}$Ge$_{46}$.\cite{Dong_PRL_2001}
Fig.\ref{fig.emd}(d)-(f) shows the average of the integrals of the HCACF calculated for 20 independent MD simulations. For the sake of clarity, only the integrated thermal conductivity in the $z$ direction for each independent case was overlaid on the corresponding average values with the final predicted thermal conductivity and theirs standard error reported in each panel. For Ge and MnGe, the calculations recover the expected isotropic value of $\kappa$ within a small error. As Mn$_5$Ge$_3$ has a hexagonal unit cell the in-plane thermal conductivity ($\kappa_{xy}$) is different from that along the $c$ axis of the crystal ($\kappa_z$). The predicted $\kappa$ of Ge, Mn$_5$Ge$_3$ and MnGe, with supercell sizes of 5$\times$5$\times$5, 5$\times$5$\times$6 and 6$\times$6$\times$6 u.c., are respectively: 58.9$\pm$1.30, 3.3$\pm$0.15(${xy}$)/4.1$\pm$0.16($z$) and 3.0$\pm$0.13 Wm$^{-1}$K$^{-1}$. The standard error for all structures in each direction is less than 5\% which is small enough to be acceptable. Besides, the lattice thermal conductivity of Mn$_5$Ge$_3$ in \textit{z} direction (4.1 Wm$^{-1}$K$^{-1}$) is larger than that in \textit{xy} direction (3.3~Wm$^{-1}$K$^{-1}$), which can be related to the anisotropy of the hexagonal structure with space group [$P6_{3}$/mcm]. %
However, the thermal conductivity anisotropy remains moderate, as it could be expected while considering phonon dispersions along \textit{xy} and \textit{z} directions, which do not exhibit significant differences in terms of acoustic group velocities and frequency range  (see Fig.~\ref{pic:eos} (b)). 

\begin{figure}
\includegraphics[width=8cm]{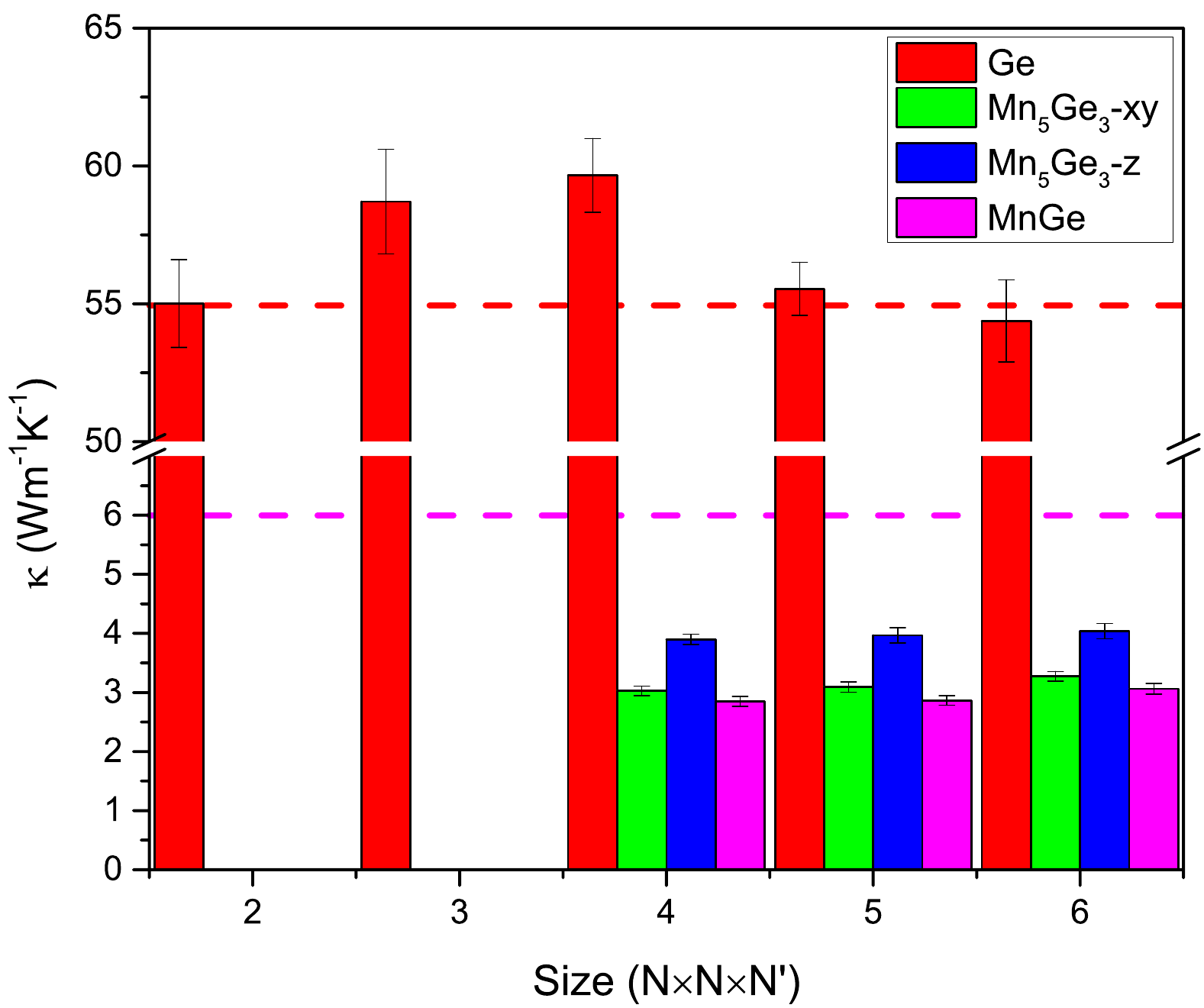}
\caption{\label{fig.kGe+kMnGe} Variation of the EMD-predicted thermal conductivity of solid Ge, Mn$_5$Ge$_3$ and MnGe at 300 K with the simulation domain size (N'). For Ge and MnGe, N=N', and for Mn$_5$Ge$_3$ N=N'-1. Each bar shows the results of 20 independent simulations with standard error overlaid on the top. The red/purple dashed horizontal line shows the thermal conductivity of solid Ge/MnGe in ferromagnetic (FM) state from DFT calculations with PBE exchange-correlation functional.}
\end{figure}
As thermal conductivity calculations by MD are particularly sensitive to size effects,\cite{Schelling:2002jl,mcgaughey2006phonon} we have performed size convergence tests for each system considered. The results obtained using supercells with increasing size are summarized in Figure~\ref{fig.kGe+kMnGe}. For the isotropic Ge and MnGe, the data are averaged in the three directions and  $\kappa$  is the result of  20 statistically independent runs with the corresponding standard error $\Delta\kappa$. For the hexagonal structure of Mn$_5$Ge$_3$, we treat  $\kappa_{xy}$ and $\kappa_{z}$ independently and report both values.
The values of the thermal conductivity of MnGe and Mn$_5$Ge$_3$ do not exhibit significant variations with size, indicating that size convergence is achieved for simulation cells of the order of 1000 atoms. 
For crystalline Ge, we observe light variations of thermal conductivity as a function of the system size, nevertheless such variations remain do not exceed  5\% of the  value obtained with the largest supercell considered (6x6x6), and the difference between the calculation with a 5$\times$5$\times$5 and with a 6$\times$6$\times$6 is well within the statistical error bars. The converged value of $\kappa$ is 54~Wm$^{-1}$K$^{-1}$, in very good agreement with experimental estimates of  $\sim$60~Wm$^{-1}$K$^{-1}$ for bulk Ge at room temperature.\cite{Glassbrenner1964thermal, asen1997thermal}
{Besides, it is seen that by using the fitted NN potential to describe the interaction between atoms in  Mn$_{\rm x}$Ge$_{1-\rm x}$ systems, the domain size has no major impact. The lattice thermal conductivities of the solid Ge, Mn$_5$Ge$_3$ and MnGe, with largest sizes of 6$\times$6$\times$6, 5$\times$5$\times$6 and 6$\times$6$\times$6 u.c. reached in our calculations, are 54.4$\pm$1.5, 3.3$\pm$0.08/4.0$\pm$0.13 (\textit{xy}/\textit{z}) and 3.1$\pm$0.09 Wm$^{-1}$K$^{-1}$, respectively.}
\textcolor{black}{
As both MnGe and Mn$_5$Ge$_3$ are metallic compounds (see spin densities of states in SI), their thermal conductivity would encompass the phononic contribution, computed in this work, and an electronic contribution ($\kappa_{el}$). The latter can be estimated through the Wiedmann-Franz law from electrical resistivity measurements: $\rho \sim 140\ \mu\Omega\cdot\rm{cm}$ for MnGe\cite{DiTusa:2014gf} and $\rho \sim 500 \ \mu\Omega\cdot\rm{cm}$ for Mn$_5$Ge$_3$.\cite{Haug:1979bx}
Assuming the ideal value for the Lorenz number, the corresponding $\kappa_{el}$ are 5.2 for MnGe and 1.5 \wmk for Mn$_5$Ge$_3$, which are of the same order as the phononic contribution.}

\begin{table}[!hbt]
\caption{Lattice thermal conductivity of Ge, ferromagnetic MnGe and Mn$_5$Ge$_3$ from equilibrium molecular dynamics with neural network potentials, compared to that obtained by DFT-BTE (Ge and MnGe) and experiments (Ge).} 
\begin{center}
\begin{tabular}{l|c|c|c}
 & Ge & MnGe & Mn$_5$Ge$_3$ \\ \hline\hline
NN+EMD & 54.4 $\pm$ 1.5 & 3.1 $\pm$ 0.09 & 4.1 / 3.3 $\pm$0.16  \\ \hline
DFT-BTE & 49.0 & 6.0 & - \\ \hline
Ref. & $\sim$60\citep{broido2007intrinsic,ward2010intrinsic,garg2011role} & - & - \\ \hline
\end{tabular}
\end{center}
\label{tab:kap}
\end{table}
In order to validate the calculations carried out with the NNP, the lattice thermal conductivity of Ge and MnGe computed with the EMD method is compared to that from DFT calculations. These results are summarized in Table~\ref{tab:kap}. For Germanium we obtain $\kappa_{NN+EMD}$=54.4$\pm$1.5 Wm$^{-1}$K$^{-1}$, $\kappa_{PBE+DFT}$=49.0 Wm$^{-1}$K$^{-1}$.  Both values are close to those reported in previous works $\sim$60 Wm$^{-1}$K$^{-1}$,\citep{broido2007intrinsic,ward2010intrinsic,garg2011role} which nevertheless use the local density approximation (LDA) for the exchange correlation functional. 
The lattice thermal conductivity of ferromagnetic MnGe from DFT-BTE calculations is 6.0 Wm$^{-1}$K$^{-1}$, which is almost twice as much as the value of 3.1$\pm$0.09 Wm$^{-1}$K$^{-1}$ from NN potential and EMD method. 
This discrepancy may probably arise from the fact that MD simulations include all order of anharmonicity, while our DFT-BTE calculations truncate the expansion of the potential to the third order, thus including only three-phonon scattering processes. In fact, recent works pointed out the importance of four-phonon scattering, especially in strongly anharmonic systems,\cite{Feng:2016ib} which leads to substantial discrepancies between MD and BTE calculations.\cite{Puligheddu:2019gr}
We stress that such large discrepancies do not necessarily stem from the complexity of the system or from the use of DFT. In fact, even with simple Lennard-Jones potentials BTE and MD results start to substantially diverge at relatively low temperature, where one would naively expect anharmonic lattice dynamics to be still a good approximation.\cite{Turney:2009bb}
Given the capability of NNP to accurately reproduce the phonon dispersion relations and the equation of state of MnGe, and the inclusion of all orders of phonon-phonon scattering in MD, we tend to consider the lower value of $\kappa$ obtained by EMD the best prediction for the yet unmeasured thermal conductivity of MnGe.   

\section{Conclusions}
\textcolor{black}{In summary, we have shown that a NNP, trained over a relatively small set of crystalline configurations, provides a satisfactory description of the structural and vibrational properties of Mn$_x$Ge$_y$ compounds over a broad range of chemical compositions. The NNP is also able to predict reasonably well the equation of state and the phonon dispersion relations of a crystalline phase, Mn$_{11}$Ge$_8$, which was not used for training, and it enables the calculation of the phonon dispersion relations of Ge/Mn$_5$Ge$_3$ heterostructures.  
In spite of numerical discrepancies between the thermal conductivity the Mn$_x$Ge$_y$ compounds computed by DFT-based anharmonic and with the NNP, our work provides the proof of principles that the NNP can be used to reliably compute the thermal conductivity of complex systems by MD across a variety of compositions and chemical environments. 
This is especially important because linear-scaling MD simulations allow one to take into account phonon scattering at all orders, which is crucial to achieve accurate predictions of $\kappa$ for complex systems with strong anharmonicity.
This study may be considered as a proof of principles of the transferability of NNPs to compute the thermal conductivity of complex materials over different compositions. This approach may be improved in several ways, for example by exploring NNPs with different structure or using different symmetry functions.\cite{Bartok:2013cs,THOMPSON2015316,Zuo:2020kp} Further efforts may be exerted to improve the construction of the training database by exploring more efficiently the configurational space, so to reduce redundancies and overfitting.\cite{Bonati:2019ih}   }

\section*{Data Availability}
{The data that supports the findings of this study are available within the article and its supplementary material.}

\section*{Supplementary Material}
Supplementary material includes the spin density of states for MnGe and Mn$_5$Ge$_3$ ferromagnetic compounds, and the parameter files of the neural network potential in the format interpreted by the RuNNer code.

\begin{acknowledgments}
The authors thank Gabriele C. Sosso and Jinming Dong for fruitful discussions. 
Financial support was provided by the European Commission FP7 FET Energy Project MERGING (Grant No. 309150), and by the ANR project MESOPHON (ANR-15-CE30-0019).
JB gratefully acknowledges a DFG Heisenberg professorship (Be3264/11-2, project number 329898176).
We  acknowledge the provision of computing facilities and support by the Rechenzentrum Garching of the Max Planck Society (MPG), the supercomputer SuperMUC at the Leibniz Rechenzentrum (project no. pr87bi), the GENCI-IDRIS High Performance Computing resources (Grant No. A0030907186, and  the ``Lorraine Universit\'{e} computation center EXPLOR''.
\end{acknowledgments}




\bibliography{biglib}

\end{document}